\documentclass[12 pt,journal,letterpaper,oneside,onecolumn]{IEEEtran}
\usepackage{graphicx} 
\usepackage{fancyhdr}
\usepackage{hyperref}
\usepackage{amssymb,amsmath,amsfonts}
\usepackage{suffix}
\usepackage{imakeidx}
\usepackage{balance}

\ifCLASSINFOpdf
\else
\fi
\begin{document}
%
\title{On the Secrecy Performance of SWIPT Receiver Architectures with Multiple Eavesdroppers}
%
%
%

\author{Furqan Jameel,
        Shurjeel Wyne,
        Syed Junaid Nawaz,
				Junaid Ahmed,\\ and
				Kanapathippillai Cumanan
\thanks{Furqan Jameel, Shurjeel Wyne, Syed Junaid Nawaz, and Junaid Ahmed are with the Department of Electrical Engineering, COMSATS Institute of Information Technology, Islamabad 45550, Pakistan.}
\thanks{Kanapathippillai Cumanan is with the Department of Electronic Engineering, University of York, Heslington, York, YO10 5DD, UK.}
}

\maketitle

\begin{abstract}
Physical layer security (PLS) has been shown to hold promise as a new paradigm for securing wireless links. In contrast with the conventional cryptographic techniques, PLS methods exploit the random fading in wireless channels to provide link security. As the channel dynamics prevent a constant rate of secure communications between the legitimate terminals, the outage probability of the achievable secrecy rate is used as a measure of the secrecy performance. This work investigates the secrecy outage probability of a simultaneous wireless information and power transfer (SWIPT) system, which  operates in the presence of multiple eavesdroppers that also have the energy harvesting capability. The loss in secrecy performance due to eavesdropper collusion, i.e., information sharing between the eavesdroppers to decode the secret message, is also analyzed. We derive closed-form expressions for the secrecy outage probability for Nakagami-$m$ fading on the links and imperfect channel estimation at the receivers. Our analysis considers different combinations of the separated and the integrated SWIPT receiver architectures at the receivers. Numerical results are provided to validate our analysis.
\end{abstract}
%
%
%
\IEEEpeerreviewmaketitle
\section{Introduction}
Simultaneous wireless information and power transfer (SWIPT) systems have spurred considerable research interest in both academia and industry \cite{krikidis2014simultaneous}. The SWIPT technique provides significant convenience to its users by efficiently utilizing the radio frequency (RF) signal for both information and power transfer \cite{huang2013simultaneous}. However, SWIPT systems require a special receiver design to support the dual capability of energy harvesting (EH) and information decoding (ID). In the literature, two broad categories of SWIPT receiver architectures have been proposed namely the separated and the integrated receiver architectures \cite{krikidis2014simultaneous}. The separated receiver architecture has dedicated separate units for ID and EH. However, this increases the complexity and cost of the receiver hardware \cite{zhou2013wireless}. In contrast, the integrated receiver architecture has a unified circuitry to perform ID and EH jointly, which reduces the hardware costs \cite{zhou2013wireless}.

Varshney et al. in \cite{varshney2008transporting} were the first to propose the transmission of information and energy simultaneously. They developed a capacity-energy function to characterize the fundamental tradeoff in performance between simultaneous information and power transfer. In \cite{grover2010shannon}, the authors extended the work of \cite{varshney2008transporting} to frequency-selective channels with additive white Gaussian noise (AWGN). It was shown in \cite{grover2010shannon} that a non-trivial tradeoff exists for information transfer versus energy transfer via power allocation. A SWIPT system under co-channel interference was studied in \cite{liu2013wireless}. The authors derived optimal designs to achieve outage-energy tradeoffs and rate-energy tradeoffs. In \cite{xiang2012robust} the authors considered the performance of a SWIPT system with imperfect channel state information (CSI) at the transmitter. Networks that employ pure wireless power transfer were studied in \cite{huang2014enabling} and \cite{lee2013opportunistic}. In \cite{huang2014enabling}, the authors studied a hybrid network that overlaid an uplink cellular network with randomly deployed power beacons, which charged mobiles wirelessly. The authors then derived the tradeoffs between different network parameters under an outage constraint on the data links.

The broadcast nature of wireless signals implies that nodes other than the intended receiver may also receive the transmitted message, which results in information leakage. Although cryptography-based techniques are conventionally used to secure transmitted information, the high computational complexity of these techniques consumes a significant amount of energy \cite{de2008energy}. Recently, physical layer security (PLS) has been proposed as an alternative for securing wireless communications by exploiting the channel characteristics such as fading, noise, and interferences \cite{Cumanan2017_access}. The secrecy performance of a cooperative network was investigated in \cite{juan2016relay, deng2017secrecy}; secrecy for interference limited networks was studied in \cite{he2009secure} and for cognitive radio networks in \cite{zou2013physical,shu2013physical,sakran2012proposed}. In\cite{Cumanan2016_JSTSP}, the authors analyzed the secrecy performance of a multicast network in which the transmitter broadcasted its information to a set of legitimate users in the presence of multiple eavesdroppers. The authors then proposed power minimization and secrecy rate maximization schemes for the considered multicasting secrecy network. The security of large-scale networks has also been characterized in terms of connectivity \cite{pinto2012secure}, coverage \cite{vilela2011wireless} and capacity \cite{koyluoglu2012secrecy}. Researchers have also considered so-called artificial noise generation techniques to reduce the signal-to-interference ratio of the eavesdropper channel while minimizing the interference to the legitimate link \cite{romero2012outage,zheng2011optimal}. The authors in \cite{Cumanan2017_TVT,liu2013destination} studied cooperative jamming, whereby a relay transmitted an interfering signal towards the eavesdropper while the source broadcasted its message. In \cite{yang2013cooperative}, secure beamforming techniques have been explored to maximize the received power at the legitimate receiver. The PLS techniques are naturally applicable to SWIPT but the design of an optimal PLS techniques for SWIPT systems is a non-trivial task since it needs to also consider the efficiency of the wireless power transfer. In general, if a power receiver is a potential eavesdropper then any increase in the information signal power to improve the power transfer efficiency may also compromise the message secrecy \cite{liu2013wireless}. Therefore, the inherent tradeoff between power efficiency and information security in a SWIPT system merits detailed examination. The authors in \cite{7904706} investigated the maximization of secrecy throughput for SWIPT systems. In particular, they considered power allocation between EH and ID to provide an optimal secure SWIPT solution. In the same work an analytical expression for the secrecy outage probability was also derived. In \cite{jameel2017secrecy}, the authors investigated the secrecy performance of a SWIPT system with the separated receiver SWIPT architecture employed at the eavesdropper and $\kappa-\mu$ faded links. In \cite{fang2015aided} the authors introduced an artificial noise-aided precoding scheme to maximize the secrecy rate. In \cite{zhang2016artificial} the authors studied the secrecy capacity of an EH orthogonal-frequency-division-multiplexing network. All the sub-carriers were allocated an identical power and the power-splitting technique was used to coordinate ID and EH. In \cite{li2014secure} the authors analyzed secure beamforming for an amplify-and-forward two-way relaying SWIPT network and proposed a zero-forcing based sub-optimal solution to maximize the secrecy of the considered network.

In the SWIPT literature most investigations have considered only the separated receiver architecture \cite{7904706,fang2015aided,zhang2016artificial,li2014secure}. Furthermore, multiple eavesdroppers when considered are often assumed to operate independently, whereas in many practical scenarios these eavesdroppers may collaborate to enhance their secret message decoding capability \cite{chen2016secrecy}. Finally, the achievable secrecy rate may degrade significantly under imperfect channel estimation at the legitimate receiver, whereas imperfect CSI at the eavesdropper can prove beneficial for the system's secrecy performance. To the best of the authors' knowledge, a comparative analysis of the secrecy performance of the separated and integrated SWIPT architectures with eavesdropper cooperation and imperfect CSI has not been performed previously. Specifically, the main contributions of the submitted work are listed as follows:
\begin{itemize}
\item We derive closed-form expressions for the secrecy outage probability with imperfect CSI knowledge at the receivers and different combinations of the separated and the integrated SWIPT architectures at the legitimate and the eavesdropping receivers. 
\item The tradeoff between secrecy performance and harvested energy is investigated.  
\item The loss in secrecy performance due to eavesdropper cooperation is analyzed and compared with the non-cooperative case.
\end{itemize}
The remainder of this paper is organized as follows. Section II presents the system model. In Section III the closed-form expressions for the outage probability are derived for different receiver architectures. Section IV provides numerical results along with relevant discussion. In Section V, some concluding remarks are given.
\section{System Model}
\begin{figure}[h]
\centering
\includegraphics[scale=.3]{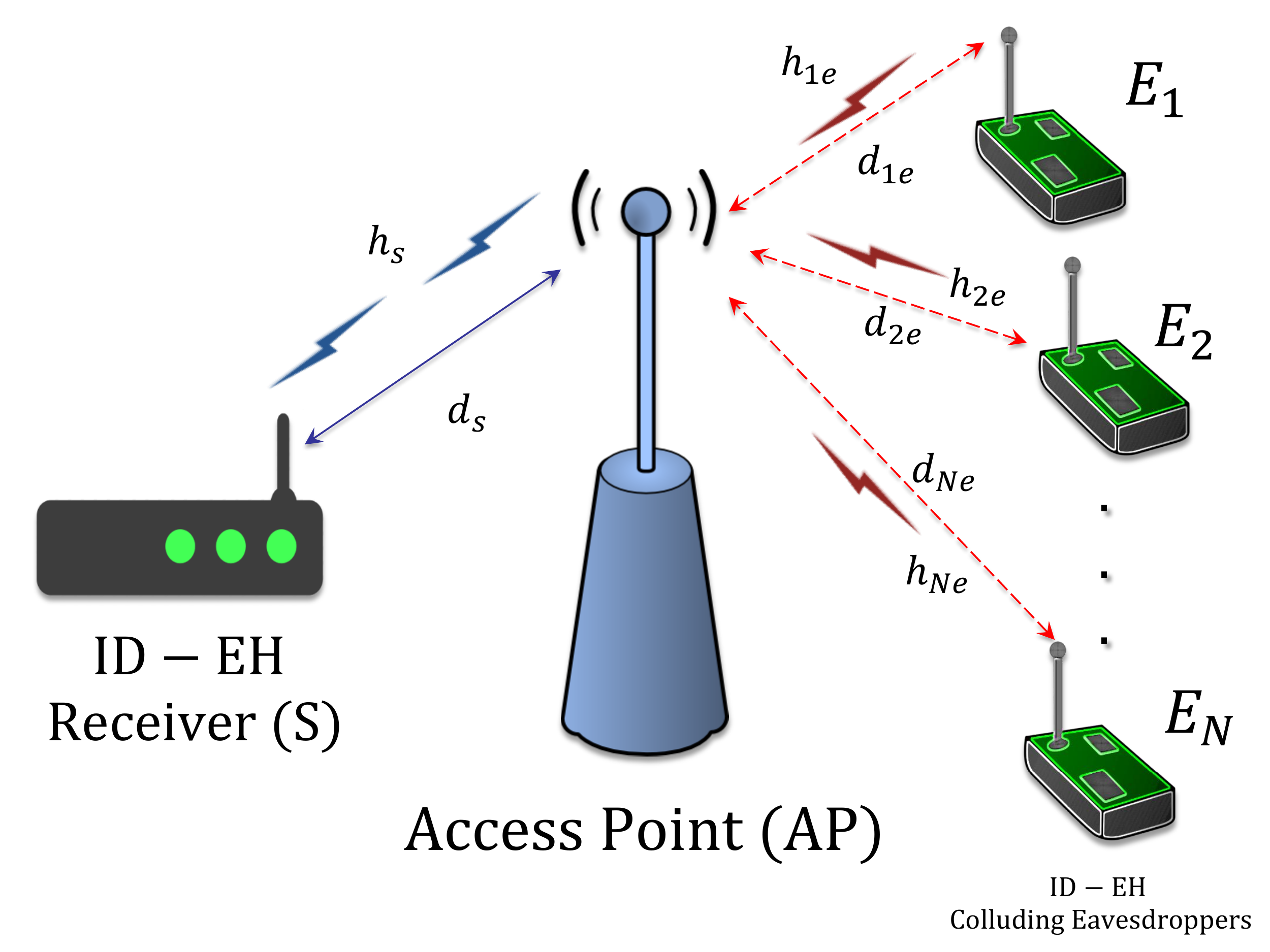}
\caption{System Model.}
\label{fig.8}
\end{figure}

\begin{figure}[h]
\centering
\includegraphics[scale=.35]{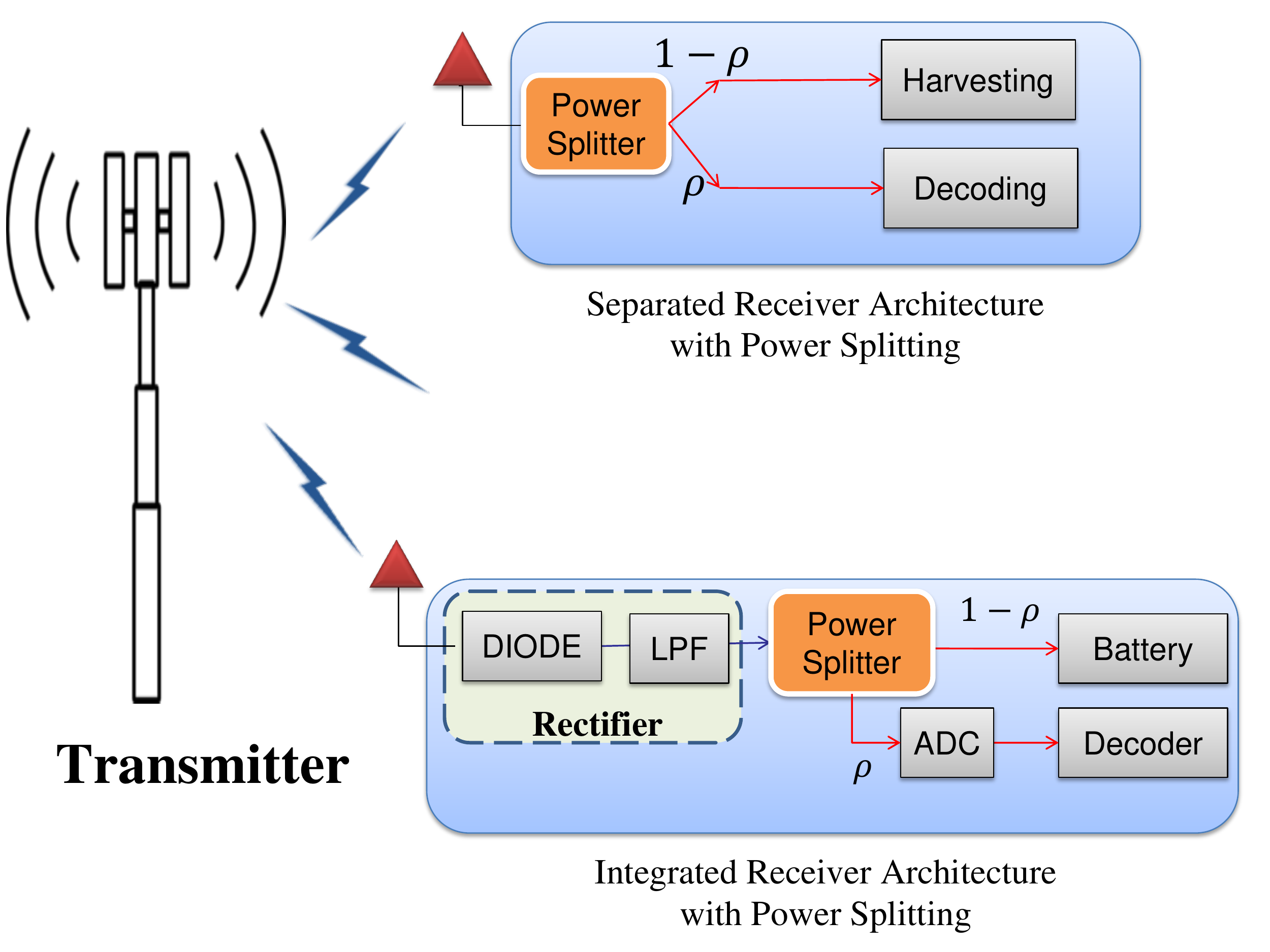}
\caption{Separated and integrated receiver architectures of SWIPT \cite{zhou2013wireless}.}
\label{fig.int}
\end{figure}

We consider the downlink of a SWIPT system as shown in Fig. \ref{fig.8} in which the Access Point (AP) transmits a secure message to the legitimate receiver S, which has simultaneous EH and ID capability. This transmission is also received by $N$ eavesdropping nodes that are admitted into the network for EH-only but exploit their SWIPT receiver architectures in an attempt to intercept the secret communication between AP and S \cite{7892953}. Since the eavesdroppers, denoted by $E=\{E_i|i=1,2,...N\}$, are also part of the network - the AP is assumed to have CSI for the main channel to node S as well as for the $N$ wiretap channels \cite{7892953}. All nodes are considered to be equipped with single antennas.\footnote{Analysis for multi-antenna nodes \cite{Zhang2016_TWC} will be reported in future work.} Our analysis considers two types of receiver architectures for both S and $E$, i.e., the conventional separated receiver and the integrated receiver architecture \cite{zhou2013wireless} shown in Fig. \ref{fig.int}. In the separated receiver, the RF signal after power-splitting (PS) is fed to separate circuitry for ID and EH, whereas in the integrated receiver PS between EH and ID takes place after the rectifier. The rectifier of the integrated receiver also down-converts the RF signal for ID, i.e., the down-conversion operation is integrated with the energy receiver in this architecture. For both receiver types, the fractional powers received for ID and EH are denoted by $0\leq\rho<1$ and $1-\rho$, respectively.

Consider that the AP transmits signal $s$ with power $P$. The signal received at S can then be written as 
\begin{align}
y_{s}=\sqrt{\frac{P}{P_{s}^{loss}}}\hat{h}_{s}s+n_{s},
\end{align}
where $\hat{h}_s$ represents the channel gain estimated by S and $n_s$ denotes the zero-mean variance $N_0$ additive white Gaussian noise (AWGN) due to the receiver electronics at S. Additionally, $P_s^{loss}=\frac{(4\pi)^2 d_s^{\Xi}}{G_t G_r \lambda_c^2}$ is the path loss, where $d_s$ denotes the distance between AP and S and $\Xi$ is the path loss exponent. Furthermore, $\lambda_c$ is the carrier wavelength and $G_t$ and $G_r$ are the antenna gains at AP and S, respectively. 

Since S employs PS architecture, the received signal is further divided into two streams for ID and EH. The signal at the information decoder of S is given as 
\begin{align}
y_{s}=\sqrt{\rho 
_{s}}\left(\sqrt{\frac{P}{P_{s}^{loss}}}\hat{h}_{s}s+n_{s}\right)+z_{s},
\label{eq_RXsignal_s}
\end{align}
where $\rho_s$ is the power splitting factor at S and $z_s$ is the signal processing noise at S, also distributed normally as $\mathcal{N}(0,\sigma_s^2)$. Since $1-\rho_s$ fraction of received power is used for energy harvesting, thus the amount of harvested energy at S, ignoring small amount of energy stored by antenna and signal processing noise, can be written as \cite{zhou2013wireless} 
\begin{align}
EH_s=\frac{\zeta_s(1-\rho_s)P|\hat{h}_{s}|^2}{P_s^{loss}},
\label{ref_eq}
\end{align}
where $\zeta_s$ represents the power conversion efficiency at S. The AP transmission is also picked up by the eavesdroppers, the signal received at the information decoder of the $i$-th eavesdropper is written as 
\begin{align}
y_{ie}=\sqrt{\rho 
_{ie}}\left(\sqrt{\frac{P}{P_{ie}^{loss}}}\hat{h}_{ie}s+n_{ie}\right)+z_{ie},
\label{hat}
\end{align}
where $\hat{h}_{ie}$ represents the channel gain estimated by the $i$-th eavesdropper. Furthermore, $n_{ie}=n_{e}$ represents the thermal noise distributed as $\mathcal{N}(0,N_0)$ and $z_{ie}=z_{e}$ is the signal processing noise distributed as $\mathcal{N}(0,\sigma_{e}^2)$, at the $i$-th eavesdropper. Here the noise statistics are assumed identical due to all eavesdroppers using the same type of hardware. For a tractable analysis, we consider $P_{ie}^{loss}=P_{e}^{loss}$ and $\rho_{ie}=\rho_e \forall i \ \in \ N$. Similar to (\ref{ref_eq}), the amount of harvested energy at the $i$-th eavesdropper can be written as \cite{zhou2013wireless}
\begin{align}
EH_{ie}=\frac{\zeta_{ie}(1-\rho_{ie})P|\hat{h}_{ie}|^2}{P_{ie}^{loss}},
\end{align}
where $\zeta_{ie}$ is the power conversion efficiency at the $i$-th eavesdropper. Moreover, without loss of generality, we consider $\zeta_{ie}= \zeta{e}$ throughout this work. Finally, the receiver nodes make an erroneous channel estimate due to their hardware impairments modeled as \cite{isukapalli2010packet,yoo2006capacity} 
\begin{align}
\hat{h}_{k}=\sqrt{1-\delta _{k}^{2}}h_{k}+\delta _{k}v,
\label{eq_CSImodel}
\end{align}
where $k\in \{s,ie\}$, ${h}_k$ represents the true channel amplitude gain. The parameter $0<\delta_k<1$ is a measure of estimation accuracy with $\delta_k = 0$ for a perfect estimate. Additionally, $v$ is a normal random variable distributed as $\mathcal{N}(0,1)$. Now by substituting (\ref{eq_CSImodel}) into (\ref{eq_RXsignal_s}) we can express the signal received at S as
\begin{align}
y_{s}=\sqrt{\rho _{s}}\left(\sqrt{\frac{P(1-\delta 
_{s}^{2})}{P_{s}^{loss}}}h_{s}s+\sqrt{\frac{P}{P_{s}^{loss}}}\delta 
_{s}vs+n_{s}\right)+z_{s},
\end{align}
and substituting (\ref{eq_CSImodel}) into (\ref{hat}) we can express the signal received at $i$-th eavesdropper as
\begin{align}
y_{ie}=\sqrt{\rho _{ie}}\left(\sqrt{\frac{P(1-\delta 
_{ie}^{2})}{P_{e}^{loss}}}h_{ie}s+\sqrt{\frac{P}{P_{e}^{loss}}}\delta 
_{ie}vs+n_{ie}\right)+z_{ie}.
\end{align}
Using the above equations, the instantaneous signal-to-noise ratio (SNR) of the main channel can be written as
\begin{align}
\chi_{s}=\frac{\rho _{s}\Omega _{s}(1-\delta _{s}^{2})}{(\Omega 
_{s}\rho _{s}\delta _{s}^{2}+\rho _{s}N_{0}+\sigma _{s}^{2})}\vert 
h_{s}\vert ^{2},
\end{align}
and the SNR for the $i$-th wiretap channel can be expressed as
\begin{align}
\chi_{ie}=\frac{\rho _{e}\Omega _{e}(1-\delta _{ie}^{2})}{(\Omega 
_{e}\rho _{s}\delta _{ie}^{2}+\rho _{e}N_{0}+\sigma _{e}^{2})}\vert 
h_{ie}\vert ^{2},
\end{align}
where $\Omega_s=P/(P_s^{loss})$ and $\Omega_e=P/(P_e^{loss})$. For subsequent analysis, $\delta_{ie}=\delta_e, \forall i \ \in \ N$ is considered. 
\section{Secrecy Outage Analysis} 
In this section closed form expressions for the secrecy outage probability are derived separately for four different cases that are based on the receiver types used at S and $E$. Specifically, $P_{out}^{Sp-Sp}$ denotes outage probability for the case of separated receiver architectures at S and $E$, $P_{out}^{Sp-In}$ denotes the outage for separated receiver at S and integrated receiver at $E$, $P_{out}^{In-Sp}$ is the outage for integrated receiver at S and separated receiver at $E$, and $P_{out}^{In-In}$ denotes outage probability for the case of integrated receivers at both S and $E$. Each of these four cases are discussed first for the non-cooperative eavesdropping scenario and later for cooperation among the eavesdroppers.
\subsection{Non-cooperative Eavesdroppers}
In this scenario, the worst-case of the eavesdropper with the maximum SNR is considered to decode the message. The instantaneous SNR of the wiretap link can be re-written as
\begin{align}
\chi _{e}=\max _{i \in N}\chi _{ie}=\frac{\rho _{e}\Omega 
_{e}(1-\delta _{e}^{2})}{(\Omega _{e}\rho _{s}\delta _{e}^{2}+\rho 
_{e}N_{0}+\sigma _{e}^{2})}\max _{i \in N} \vert h_{ie}\vert ^{2}.
\end{align}
where $\chi_{ie}$ is Gamma distributed \cite{karagiannidis2003multivariate} with probability density function (PDF) expressed as\\%
\vspace{0.5cm}
 $f_{\chi_{ie}}(\gamma _{ie})=\left[ \frac{m_{ie}(\Omega _{ie}\rho _{ie}\delta _{ie}^{2}+\rho {ie}N_{0}+\sigma _{ie}^{2})}{\rho _{ie}(1-\delta _{ie}^{2})\bar{\gamma }_{ie}}\right] ^{m_{ie}} \times \exp\left(-\frac{m_{ie}(\Omega _{ie}\rho _{ie}\delta _{ie}^{2}+\rho _{ie}N_{0}+\sigma _{ie}^{2})\gamma _{ie}}{\rho _{ie}(1-\delta _{ie}^{2})\bar{\gamma }_{ie}}\right) \times \frac{(\gamma _{ie})^{m_{ie}-1}}{\Gamma(m_{ie})}$.\\%
\vspace{0.5cm}
 Then, the cumulative distribution function (CDF) for the instantaneous SNR of the wiretap link (i.e. random variable $\chi _{e}$ falling below an arbitrary value $\gamma_{e}$), is given as 
\begin{align}
F_{\chi _{e}}(\gamma _{e})&=\Pr(\chi _{e}<\gamma _{e}).
\end{align}
Now using statistical independence of the wiretap channels and the CDF of a Gamma random variable \cite{morsi2015multiuser}, we obtain
\begin{align}
F_{\chi _{e}}(\gamma _{e})&=\Pr(\chi _{1e}<\gamma _{e},\chi _{2e}<\gamma _{e},\ldots, \chi_{Ne}<\gamma _{e}),\nonumber \\
&=\biggl[ 1-\exp\left(-\frac{m_{e}(\Omega _{e}\rho 
_{e}\delta _{e}^{2}+\rho _{e}N_{0}+\sigma _{e}^{2})\gamma _{e}}{\rho 
_{e}(1-\delta _{e}^{2})\bar{\gamma}_{e}}\right) \nonumber \\
&\times \sum_{r=0}^{m_{e}-1}\frac{1}{r!}\left[ \frac{m_{e}(\Omega_{e}\rho _{e}\delta _{e}^{2}+\rho _{e}N_{0}+\sigma _{e}^{2})\gamma 
_{e}}{\rho _{e}(1-\delta _{e}^{2})\bar{\gamma }_{e}}\right]^{r} \biggr]^{N}, 
\label{Nak_cdf_e}
\end{align}

The corresponding PDF can be written as
\begin{align}
f_{\chi _{e}}(\gamma _{e})&=\frac{dF_{\chi_{e}}(\gamma 
_{e})}{d\gamma _{e}} \nonumber \\
&=\frac{N(\gamma _{e})^{m_{e}-1}}{\Gamma 
(m_{e})}\left[\frac{m_{e}(\Omega _{e}\rho _{e}\delta _{e}^{2}+\rho 
_{e}N_{0}+\sigma _{e}^{2})}{\rho _{e}(1-\delta _{e}^{2})\bar{\gamma 
}_{e}}\right]^{m_{e}} \nonumber \\
& \times \biggl[ 1-\exp\left(-\frac{m_{e}(\Omega_{e}\rho _{e}\delta _{e}^{2}+\rho _{e}N_{0}+\sigma _{e}^{2})\gamma _{e}}{\rho _{e}(1-\delta _{e}^{2})\bar{\gamma}_{e}}\right) \nonumber \\
&\times \sum_{r=0}^{m_{e}-1}\frac{1}{r!}\left\{ \frac{m_{e}(\Omega_{e}\rho _{e}\delta _{e}^{2}+\rho _{e}N_{0}+\sigma _{e}^{2})\gamma_{e}}{\rho _{e}(1-\delta _{e}^{2})\bar{\gamma }_{e}}\right\}^{r}\biggr] ^{N-1} \nonumber \\
&\times \exp \left(-\frac{m_{e}(\Omega _{e}\rho _{e}\delta 
_{e}^{2}+\rho _{e}N_{0}+\sigma _{e}^{2})\gamma _{e}}{\rho _{e}(1-\delta 
_{e}^{2})\bar{\gamma }_{e}}\right), 
\label{Nak_pdf_e}
\end{align}
where $\bar{\gamma}_{e}=\Omega_e\mathbb{E}\{\max_{i \in N}|h_{ie}|^2\}$ represents the average SNR of the wiretap link and $m_e$ is the Nakagami-m fading severity parameter for the wiretap link.

The PDF of the instantaneous SNR of the main link can be obtained as \cite{karagiannidis2003multivariate}
\begin{align}
f_{\chi_{s}}(\gamma _{s})&=\left[ \frac{m_{s}(\Omega _{s}\rho _{s}\delta _{s}^{2}+\rho 
_{s}N_{0}+\sigma _{s}^{2})}{\rho _{s}(1-\delta _{s}^{2})\bar{\gamma 
}_{s}}\right] ^{m_{s}} \nonumber \\
&\times \frac{(\gamma _{s})^{m_{s}-1}\exp\left(-\frac{m_{s}(\Omega _{s}\rho _{s}\delta 
_{s}^{2}+\rho _{s}N_{0}+\sigma _{s}^{2})\gamma _{s}}{\rho _{s}(1-\delta 
_{s}^{2})\bar{\gamma }_{s}}\right)}{\Gamma(m_{s})}.
\label{Nak_pdf_s}
\end{align}
The corresponding CDF is given as \cite{karagiannidis2003multivariate}  
\begin{align}
F_{\chi _{s}}(\gamma _{s})&=1-\exp\left(-\frac{m_{s}(\Omega _{s}\rho 
_{s}\delta _{s}^{2}+\rho _{s}N_{0}+\sigma _{s}^{2})\gamma _{s}}{\rho 
_{s}(1-\delta _{s}^{2})\bar{\gamma}_{s}}\right)\sum_{r=0}^{m_{s}-1}\frac{1}{r!} \nonumber \\
&\times \left[ \frac{m_{s}(\Omega_{s}\rho _{s}\delta _{s}^{2}+\rho _{s}N_{0}+\sigma _{s}^{2})\gamma 
_{s}}{\rho _{s}(1-\delta _{s}^{2})\bar{\gamma }_{s}}\right]^{r},
\label{Nak_cdf_s}
\end{align}
where $\bar{\gamma}_{s}=\Omega_s\mathbb{E}\{|h_s|^2\}$ is the average SNR of the main link and $m_s$ represents the Nakagami-m fading severity parameter for the main link. 
\subsubsection{Separated Receivers at S and $E$}
The achievable rates for the main and wiretap links can be written as $C_{s}=\log _{2}(1+\chi _{s})$ and $C_{e}=\log _{2}(1+\chi _{e})$, respectively \cite{zhou2013wireless}. The achievable secrecy rate $C_{sec}$ is defined as the non-negative difference between the achievable rates of the main channel and wiretap channel, which is expressed as $C_{sec }=[C_{s}-C_{e}]^{+}$. A secrecy outage event occurs when $C_{sec}$ falls below some target rate $R_s > 0$ \cite{jameel2017secure,nguyen2017secure}. The secrecy outage probability is then written as
\begin{align}
P_{out}^{Sp-Sp}&=\Pr(C_{sec }<R_{s})\nonumber\\
&=\int_{0}^{\infty}{\int_{0}^{2^{R_{s}}(1+\gamma_{e})-1}{f_{\chi_{s}}(\gamma_{s})}f_{\chi_{e}}(\gamma_{e})}d\gamma_{s}d\gamma_{e},\nonumber\\
&=\int_{0}^{\infty }{F_{\chi_{s}}(2^{R_{s}}(1+\gamma_{e})-1)f_{\chi_{e}}(\gamma_{e})}d\gamma_{e}.
\label{eq_1}
\end{align}
Now using (\ref{Nak_pdf_e}) and (\ref{Nak_pdf_s}) in (\ref{eq_1}) and with the help of \cite[(8.352.4)]{gradshteyn2014table}, we obtain%
\begin{align}%
P_{out}^{Sp-Sp}&=\frac{N}{\Gamma (m_{e})}\left[\frac{m_{e}(\Omega_{e}\rho_{e}\delta_{e}^{2}+\rho_{e}N_{0}+\sigma_{e}^{2})}{\rho_{e}(1-\delta_{e}^{2})\bar{\gamma}_{e}}\right] 
^{m_{e}}\nonumber\\
&\times\sum_{w=0}^{N-1}{\biggl(\begin{matrix}
\ N-1&\\
\ w&\\
\end{matrix}
\biggr)\frac{(-1)^{w}}{\Gamma (m_{e})\Gamma (m_{s})}}\times \mathcal{M}(\Psi _{1},\Psi _{2}),
\end{align}%
where%
\begin{align}%
\mathcal{M}(a,b)=&\int_{0}^{\infty}{(\gamma_{e})^{m_{e}-1}\exp(-m_{e}a)\Gamma(m_{e},m_{e}a)^{w}}\nonumber\\
         &\ \ \ \ \ \ \times\Gamma(m_{s},m_{s}b)d\gamma_{e},\nonumber\\
\Psi_{1}=&\frac{(\Omega_{e}\rho_{e}\delta_{e}^{2}+\rho_{e}N_{0}+\sigma_{e}^{2})\gamma_{e}}{\rho_{e}(1-\delta_{e}^{2})\bar{\gamma}_{e}},\nonumber\\
\Psi_{2}=&\frac{(\Omega_{s}\rho_{s}\delta_{s}^{2}+\rho_{s}N_{0}+\sigma_{s}^{2})(2^{R_{s}}(1+\gamma_{e})-1)}{\rho_{s}(1-\delta_{s}^{2})\bar{\gamma}_{s}}.\nonumber
\end{align}
Furthermore, $\Gamma(.,.)$ is the upper incomplete Gamma function and $\Gamma(.)$ is the Gamma function \cite{gradshteyn2014table}. The function $\mathcal{M}(a,b)$ can be readily evaluated using any computational software.
\subsubsection{Separated Receiver at S and Integrated Receiver at $E$}
In this case the achievable rate for the main link is $C_{s}=\log_{2}(1+\chi_{s})$. On the wiretap link, the integrated receiver's ID channel can be modeled as a free-space optical intensity channel \cite{zhou2013wireless}. The asymptotic high-SNR achievable rate for this channel is expressed as $C_{e}=\log_{2}(\chi_{e})+\frac{1}{2}\log_{2}\frac{e}{2\pi}$, assuming that the signal processing noise dominates the antenna noise \cite{zhou2013wireless,Lapidoth2009}. Then using the approach of (\ref{eq_1}), we obtain
\begin{align}
P_{out}^{Sp-In}=\int_{0}^{\infty }{F_{\chi_{s}}(2^{R_{s}}\gamma_{e}C-1)f_{\chi_{e}}(\gamma_{e})}d\gamma_{e},
\label{Nak_sp_in}
\end{align}
where $C=\sqrt{\frac{e}{2 \pi}}$. Substituting (\ref{Nak_pdf_e}) and (\ref{Nak_pdf_s}) in (\ref{Nak_sp_in}) and using \cite[(8.352.4)]{gradshteyn2014table}, we get
\begin{align}
P_{out}^{Sp-In}&=\frac{N}{\Gamma (m_{e})}\left[ \frac{m_{e}(\Omega 
_{e}\rho _{e}\delta _{e}^{2}+\rho _{e}N_{0}+\sigma _{e}^{2})}{\rho 
_{e}(1-\delta _{e}^{2})\bar{\gamma }_{e}}\right] 
^{m_{e}} \nonumber \\
&\times \sum_{w=0}^{N-1}{\biggl(\begin{matrix}
\ N-1&\\
\ w&\\
\end{matrix}
\biggr)\frac{(-1)^{w}}{\Gamma (m_{e})\Gamma (m_{s})}} \times \mathcal{M}(\Psi _{1},\Psi _{3}).
\end{align}
where $\Psi _{3}=\frac{(\Omega _{s}\rho _{s}\delta _{s}^{2}+\rho _{s}N_{0}+\sigma _{s}^{2})(2^{R_{s}}\gamma _{e}C-1)}{\rho _{s}(1-\delta 
_{s}^{2})\bar{\gamma }_{s}}$.
\subsubsection{Integrated Receiver at S and Separated Receiver at $E$}
In this case, the main link has an asymptotic achievable rate of $C_{s}=\log_{2}(\chi_{s})+\frac{1}{2}\log_{2}\frac{e}{2\pi}$ \cite{zhou2013wireless,Lapidoth2009}, whereas the achievable rate for the wiretapper is $C_{e}=\log_{2}(1+\chi_{e})$. Then using a similar approach to (\ref{eq_1}) and after some manipulations, the outage probability is given as
\begin{align}
P_{out}^{In-Sp}=1-\int_{\frac{2^{R_{s}}}{C}}^{\infty }{F_{\chi_{e}}\left(\frac{\gamma_{s}C}{2^{R_{s}}}-1\right)}f_{\chi_{s}}(\gamma_{s})d\gamma_{s}.
\label{eq_IN_SP}
\end{align}
Substituting (\ref{Nak_cdf_e}) and (\ref{Nak_pdf_s}) in (\ref{eq_IN_SP}) and using the binomial theorem, we get
\begin{align}
P_{out}^{In-Sp}&=1-\sum_{z=0}^{N}{\biggl(\begin{matrix}
\ N & \\
\ z & \\
\end{matrix}
\biggr)\frac{(-1)^{z}}{\Gamma (m_{s})\Gamma (m_{e})}} \nonumber \\
&\times \left[ 
\frac{m_{s}(\Omega _{s}\rho _{s}\delta _{s}^{2}+\rho _{s}N_{0}+\sigma 
_{s}^{2})}{\rho _{s}(1-\delta _{s}^{2})\bar{\gamma }_{s}}\right] ^{m_{s}} \mathcal{T}(\Psi _{4},\Psi _{5}),
\end{align}
where $\mathcal{T}(a,b)=\int_{2^{R_{s}}/C}^{\infty}{\Gamma(m_{e},m_{e}a)^{z}(\gamma_{s})^{m_{s}-1}\exp(-m_{s}b)d\gamma_{s}}$ involves a single integral and can be readily evaluated in any computational software. Furthermore, $\Psi_{4}=\frac{(\Omega_{e}\rho_{e}\delta_{e}^{2}+\rho_{e}N_{0}+\sigma_{e}^{2})(\frac{\gamma_{s}C}{2^{R_{s}}}-1)}{\rho_{e}(1-\delta_{e}^{2})\bar{\gamma}_{e}}$ and $\Psi_{5}=\frac{(\Omega_{s}\rho_{s}\delta_{s}^{2}+\rho_{s}N_{0}+\sigma_{s}^{2})\gamma_{s}}{\rho_{s}(1-\delta_{s}^{2})\bar{\gamma}_{s}}$.
\subsubsection{Integrated Receivers at S and $E$}
In this case the main and wiretap links have asymptotic achievable rates of $C_{s}=\log_{2}(\chi_{s})+\frac{1}{2}\log_{2}\frac{e}{2\pi}$ and $C_{e}=\log_{2}(\chi_{e})+\frac{1}{2}\log_{2}\frac{e}{2\pi}$, respectively \cite{zhou2013wireless}. Then using the same approach as that for deriving (\ref{eq_1}), we obtain
\begin{align}
P_{out}^{In-In}=1-\int_{2^{R_{s}}}^{\infty }{F_{\chi_{e}}\left(\frac{\gamma _{s}}{2^{R_{s}}}\right)}f_{\chi_{s}}(\gamma _{s})d\gamma 
_{s}.
\label{eq_IN_IN}
\end{align}
Replacing (\ref{Nak_cdf_e}) and (\ref{Nak_pdf_s}) in (\ref{eq_IN_IN}) and after some algebraic manipulations, we obtain
\begin{align}
P_{out}^{In-In}&=1-\sum_{z=0}^{N}{\biggl(\begin{matrix}
\ N & \\
\ z & \\
\end{matrix}
\biggr)\frac{(-1)^{z}}{\Gamma (m_{s})\Gamma (m_{e})}} \nonumber \\
&\times \left[\frac{m_{s}(\Omega _{s}\rho _{s}\delta _{s}^{2}+\rho _{s}N_{0}+\sigma 
_{s}^{2})}{\rho _{s}(1-\delta _{s}^{2})\bar{\gamma }_{s}}\right] ^{m_{s}}\mathcal{T}(\Psi _{6},\Psi _{5}),
\end{align}
where $\Psi _{6}=\frac{(\Omega _{e}\rho _{e}\delta _{e}^{2}+\rho _{e}N_{0}+\sigma _{e}^{2})\frac{\gamma _{s}}{2^{R_{s}}}}{\rho 
_{e}(1-\delta _{e}^{2})\bar{\gamma }_{e}}$.
\subsection{Cooperative Eavesdroppers}
For the case of cooperative eavesdropping, the $N$ eavesdroppers share information to form a virtual antenna array for receive beamforming such that a single-input multiple-output (SIMO) channel exists between the AP and the eavesdroppers \cite{pinto2012secure}. The combined message ensures the maximum achievable rate of the wiretap link. In this case the instantaneous SNR of the combined wiretap signal can be written as
\begin{align}
\chi_e=\sum^{N}_{i=1}{\chi_{ie}}.
\end{align}
The PDF of $\chi_e$ can be written as \cite{aalo1995performance}
\begin{align}
f_{\chi _{e}}(\gamma _{e})&=\left(\frac{m_{e}(\Omega _{e}\rho _{e}\delta _{e}^{2}+\rho _{e}N_{0}+\sigma _{e}^{2})}{\rho _{e}(1-\delta 
_{e}^{2})\bar{\gamma }_{e}}\right)^{Nm_{e}}\frac{\gamma 
_{e}^{Nm_{e}-1}}{\Gamma (Nm_{e})}\nonumber\\
&\times \exp(-\frac{m_{e}(\Omega _{e}\rho _{e}\delta _{e}^{2}+\rho _{e}N_{0}+\sigma _{e}^{2})}{\rho _{e}(1-\delta_{e}^{2})\bar{\gamma }_{e}}\bar{\gamma }_{e}).
\label{Nak_pdf_e_cop}
\end{align}
The CDF of the sum of independent, identically-distributed Gamma random variables is expressed as \cite{morsi2015multiuser} 
\begin{align}
F_{\chi _{e}}(\gamma _{e})&=1-{\exp\left(-\frac{m_{e}(\Omega 
_{e}\rho _{e}\delta _{e}^{2}+\rho _{e}N_{0}+\sigma _{e}^{2})\gamma 
_{e}}{\rho _{e}(1-\delta _{e}^{2})\bar{\gamma 
}_{e}}\right)} \nonumber \\
& \times \sum_{r=0}^{Nm_{e}-1}{\frac{1}{r!}}\left(-\frac{m_{e}(\Omega _{e}\rho_{e}\delta _{e}^{2}+\rho _{e}N_{0}+\sigma _{e}^{2})\gamma _{e}}{\rho_{e}(1-\delta _{e}^{2})\bar{\gamma }_{e}}\right)^{r}.
\label{Nak_cdf_e_cop}
\end{align}
\subsubsection{Separated Receivers at S and $E$}
Using (\ref{Nak_pdf_e_cop}) and (\ref{Nak_pdf_s}) in (\ref{eq_1}) and with the help of \cite[(8.352.4)]{gradshteyn2014table}, the secrecy outage probability for this case is expressed as 
\begin{align}
P_{out}^{Sp-Sp}=1-\left(\frac{m_{e}(\Omega _{e}\rho _{e}\delta _{e}^{2}+\rho 
_{e}N_{0}+\sigma _{e}^{2})}{\rho _{e}(1-\delta _{e}^{2})\bar{\gamma 
}_{e}}\right)^{Nm_{e}}\mathcal{U}(\Psi _{1},\Psi _{2}),
\end{align}
where $\mathcal{U}(a,b)=\int_{0}^{\infty }{\frac{\gamma _{e}^{Nm_{e}-1}}{\Gamma (Nm_{e})}\exp(-m_{e}a)}\frac{\Gamma (Nm_{e},m_{s}b)}{\Gamma 
(Nm_{e})}d\gamma _{e}$.
\subsubsection{Separated Receiver at S, Integrated Receiver at $E$}
Substituting (\ref{Nak_pdf_e_cop}) and (\ref{Nak_pdf_s}) into (\ref{Nak_sp_in}), the secrecy outage probability for this case is expressed as
\begin{align}
&P_{out}^{Sp-In}=1-\int_{0}^{\infty }\left(\frac{m_{e}(\Omega _{e}\rho _{e}\delta_{e}^{2}+\rho _{e}N_{0}+\sigma _{e}^{2})}{\rho _{e}(1-\delta 
_{e}^{2})\bar{\gamma }_{e}}\right)^{Nm_{e}} \nonumber \\
&\times \frac{\gamma_{e}^{Nm_{e}-1}}{\Gamma (Nm_{e})}\exp\left(-\frac{m_{e}(\Omega _{e}\rho_{e}\delta _{e}^{2}+\rho _{e}N_{0}+\sigma _{e}^{2})}{\rho _{e}(1-\delta_{e}^{2})\bar{\gamma }_{e}}\bar{\gamma }_{e}\right) \nonumber \\
&\times \exp\left(-\frac{m_{s}(\Omega _{s}\rho _{s}\delta _{s}^{2}+\rho 
_{s}N_{0}+\sigma _{s}^{2})(2^{R_{s}}\gamma _{e}C-1)}{\rho _{s}(1-\delta 
_{s}^{2})\bar{\gamma }_{s}}\right) \nonumber \\
&\times \sum_{r=0}^{m_{s}-1}{\frac{1}{r!}\left[ \frac{m_{s}(\Omega _{s}\rho _{s}\delta _{s}^{2}+\rho _{s}N_{0}+\sigma _{s}^{2})(2^{R_{s}}\gamma _{e}C-1)}{\rho _{s}(1-\delta 
_{s}^{2})\bar{\gamma }_{s}}\right]}^{r}d\gamma _{e}.
\end{align}
After some simplifications and using \cite[(8.352.4)]{gradshteyn2014table}, the secrecy outage probability is expressed as
\begin{align}
P_{out}^{Sp-In}=1-\left(\frac{m_{e}(\Omega _{e}\rho _{e}\delta _{e}^{2}+\rho 
_{e}N_{0}+\sigma _{e}^{2})}{\rho _{e}(1-\delta _{e}^{2})\bar{\gamma 
}_{e}}\right)^{Nm_{e}}\mathcal{U}(\Psi _{1},\Psi _{3}).
\end{align}
\subsubsection{Integrated Receiver at S and Separated Receiver at $E$}
Substituting (\ref{Nak_cdf_e_cop}) and (\ref{Nak_pdf_s}) in (\ref{eq_IN_SP}), the secrecy outage probability for this case is expressed as
\begin{align}
P_{out}^{In-Sp}&=1-\frac{\Gamma (m_{s},\frac{m_{s}(\Omega _{s}\rho _{s}\delta _{s}^{2}+\rho _{s}N_{0}+\sigma _{s}^{2})\gamma _{s}}{\rho _{s}(1-\delta_{s}^{2})\bar{\gamma }_{s}})}{\Gamma 
(m_{s})} \nonumber \\
&-\int_{\frac{2^{R_{s}}}{C}}^{\infty }\exp\biggl(-\frac{m_{e}(\Omega 
_{e}\rho _{e}\delta _{e}^{2}+\rho _{e}N_{0}+\sigma _{e}^{2})\gamma 
_{s}C}{\rho _{e}(1-\delta _{e}^{2})\bar{\gamma 
}_{e}2^{R_{s}}} \nonumber \\
&+\frac{m_{e}(\Omega _{e}\rho _{e}\delta _{e}^{2}+\rho 
_{e}N_{0}+\sigma _{e}^{2})}{\rho _{e}(1-\delta _{e}^{2})\bar{\gamma 
}_{e}2^{R_{s}}}\biggr)\times\sum_{r=0}^{Nm_{e}-1}{\frac{1}{r!}} \nonumber \\
&\times \biggl(-\frac{m_{e}(\Omega _{e}\rho _{e}\delta _{e}^{2}+\rho _{e}N_{0}+\sigma _{e}^{2})\gamma _{s}C}{\rho _{e}(1-\delta _{e}^{2})\bar{\gamma }_{e}2^{R_{s}}} \nonumber \\
&+\frac{m_{e}(\Omega _{e}\rho _{e}\delta _{e}^{2}+\rho _{e}N_{0}+\sigma _{e}^{2})}{\rho _{e}(1-\delta _{e}^{2})\bar{\gamma }_{e}2^{R_{s}}}\biggr)^{r} \frac{(\gamma _{s})^{m_{s}-1}}{\Gamma 
(m_{s})}\nonumber \\
&\times \left[ \frac{m_{s}(\Omega _{s}\rho _{s}\delta _{s}^{2}+\rho 
_{s}N_{0}+\sigma _{s}^{2})}{\rho _{s}(1-\delta _{s}^{2})\bar{\gamma 
}_{s}}\right] ^{m_{s}} \nonumber \\
&\times \exp\left(-\frac{m_{s}(\Omega _{s}\rho _{s}\delta _{s}^{2}+\rho 
_{s}N_{0}+\sigma _{s}^{2})\gamma _{s}}{\rho _{s}(1-\delta 
_{s}^{2})\bar{\gamma }_{s}}\right)d\gamma _{s}.
\end{align}
After some algebraic manipulations, we obtain
\begin{align}
P_{out}^{In-Sp}&=1-\frac{\Gamma (m_{s},\frac{m_{s}(\Omega _{s}\rho _{s}\delta 
_{s}^{2}+\rho _{s}N_{0}+\sigma _{s}^{2})\gamma _{s}}{\rho _{s}(1-\delta 
_{s}^{2})\bar{\gamma }_{s}})}{\Gamma (m_{s})} \nonumber \\
&-\frac{\lbrack 
\frac{m_{s}(\Omega _{s}\rho _{s}\delta _{s}^{2}+\rho _{s}N_{0}+\sigma 
_{s}^{2})}{\rho _{s}(1-\delta _{s}^{2})\bar{\gamma }_{s}}\rbrack 
^{m_{s}}}{\Gamma (m_{s})}\mathcal{V}(\Psi _{4},\Psi _{5}),
\label{neww}
\end{align}
where $\mathcal{V}(a,b)=\int_{\frac{2^{R_{s}}}{C}}^{\infty }{\frac{(\gamma _{s})^{m_{s}-1}\exp(-m_{s}a)\Gamma 
(Nm_{s},m_{e}b)}{\Gamma (Nm_{s})}d\gamma _{s}}$.
\subsubsection{Integrated Receivers at S and $E$}
Replacing (\ref{Nak_cdf_e_cop}) and (\ref{Nak_pdf_s}) in (\ref{eq_IN_IN}) and using a similar approach as for the derivation of \ref{neww}, the secrecy outage probability for this case is expressed as
\begin{align}
P_{out}^{In-In}&=1-\frac{\Gamma (m_{s},\frac{m_{s}(\Omega _{s}\rho _{s}\delta 
_{s}^{2}+\rho _{s}N_{0}+\sigma _{s}^{2})\gamma _{s}}{\rho _{s}(1-\delta 
_{s}^{2})\bar{\gamma }_{s}})}{\Gamma (m_{s})} \nonumber \\
&-\frac{\lbrack \frac{m_{s}(\Omega _{s}\rho _{s}\delta _{s}^{2}+\rho _{s}N_{0}+\sigma _{s}^{2})}{\rho _{s}(1-\delta _{s}^{2})\bar{\gamma }_{s}}\rbrack^{m_{s}}}{\Gamma (m_{s})}\mathcal{V}(\Psi _{6},\Psi _{5}).
\end{align}
\section{Numerical Results and Discussion}
We now provide some numerical results to validate the analytical expressions derived in Section III. The system parameters provided in Table \ref{tab} are used for result generation, unless stated otherwise.  
\begin{table}[hpt]
\centering
\begin{tabular}{|c|l|c|}
\hline
\textbf{S No.} & \textbf{Simulation Parameter}          & \textbf{Value} \\ \hline
1.             & Channel Realizations                   & $10^5$         			\\ \hline
2.             & Antenna Noise Variance $N_0$           & 0.1 dB        			\\ \hline
3.             & Signal Processing Noise Variance $\sigma_s^2=\sigma_e^2$ & 0 dB\\ \hline
4.             & Target Secrecy Rate $R_s$              & 1 bit/sec/Hz              \\ \hline
5.             & Main Link Power $\Omega_s$             & 30 dB              \\ \hline
6.             & Wiretap Link Power $\Omega_e$          & 10 dB             \\ \hline
7.             & Nakagami-$m$ shape factor $m_s=m_e$    & 2             \\ \hline
8.             & Power splitting factor $\rho_s=\rho_e$ & 0.8             \\ \hline
9.             & Channel estimation accuracy $\delta_s=\delta_e$   & 0.2             \\ \hline
10.            & No. of eavesdroppers $N$               & 5              \\ \hline
\end{tabular}
\vspace{2.5mm}
\caption{Simulation Parameters.}
\label{tab}
\end{table}
\begin{figure*}
  \centering
  \begin{tabular}[htb]{cc}
    \includegraphics[width=.45\textwidth]{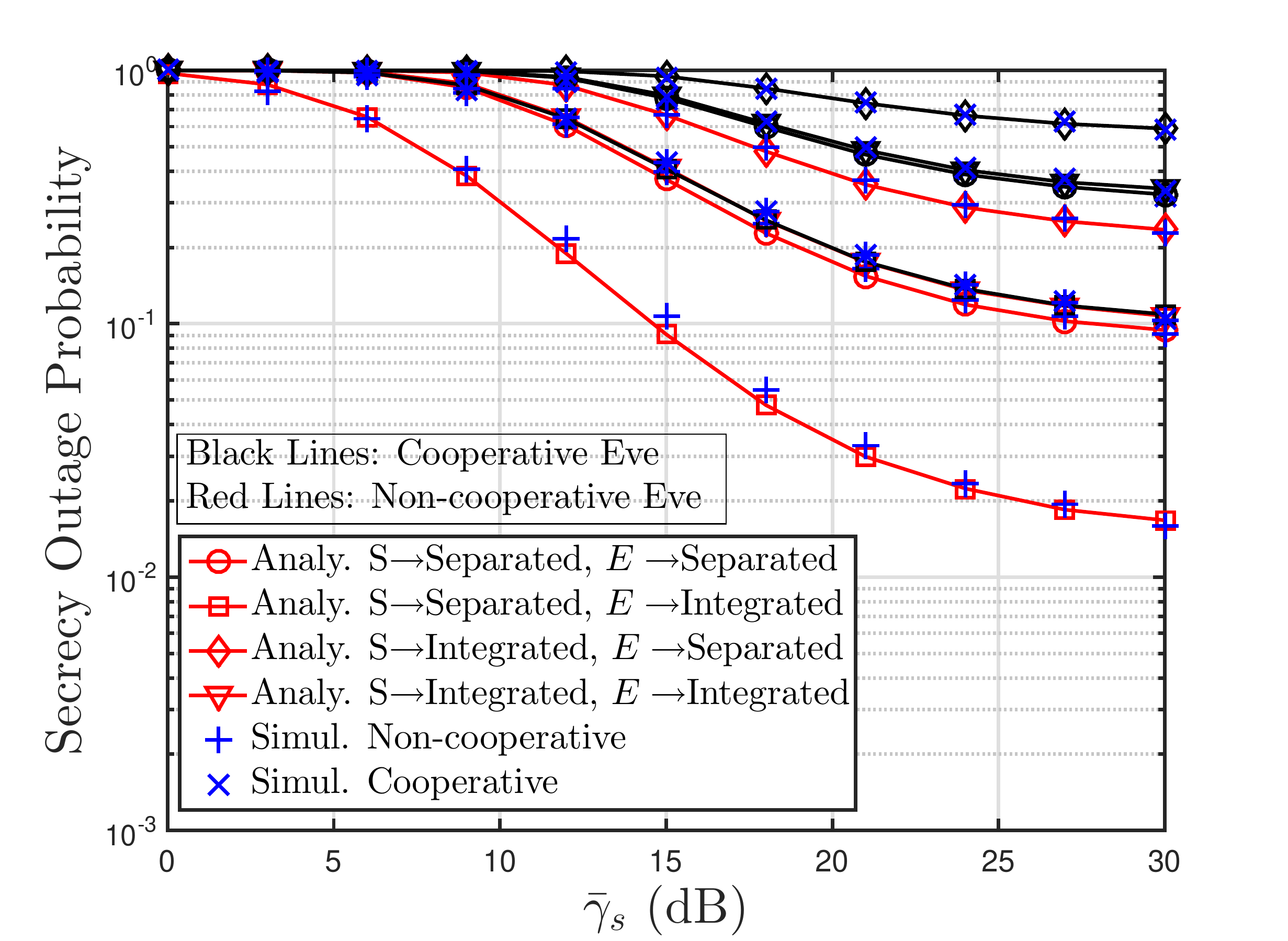}&
    \includegraphics[width=.45\textwidth]{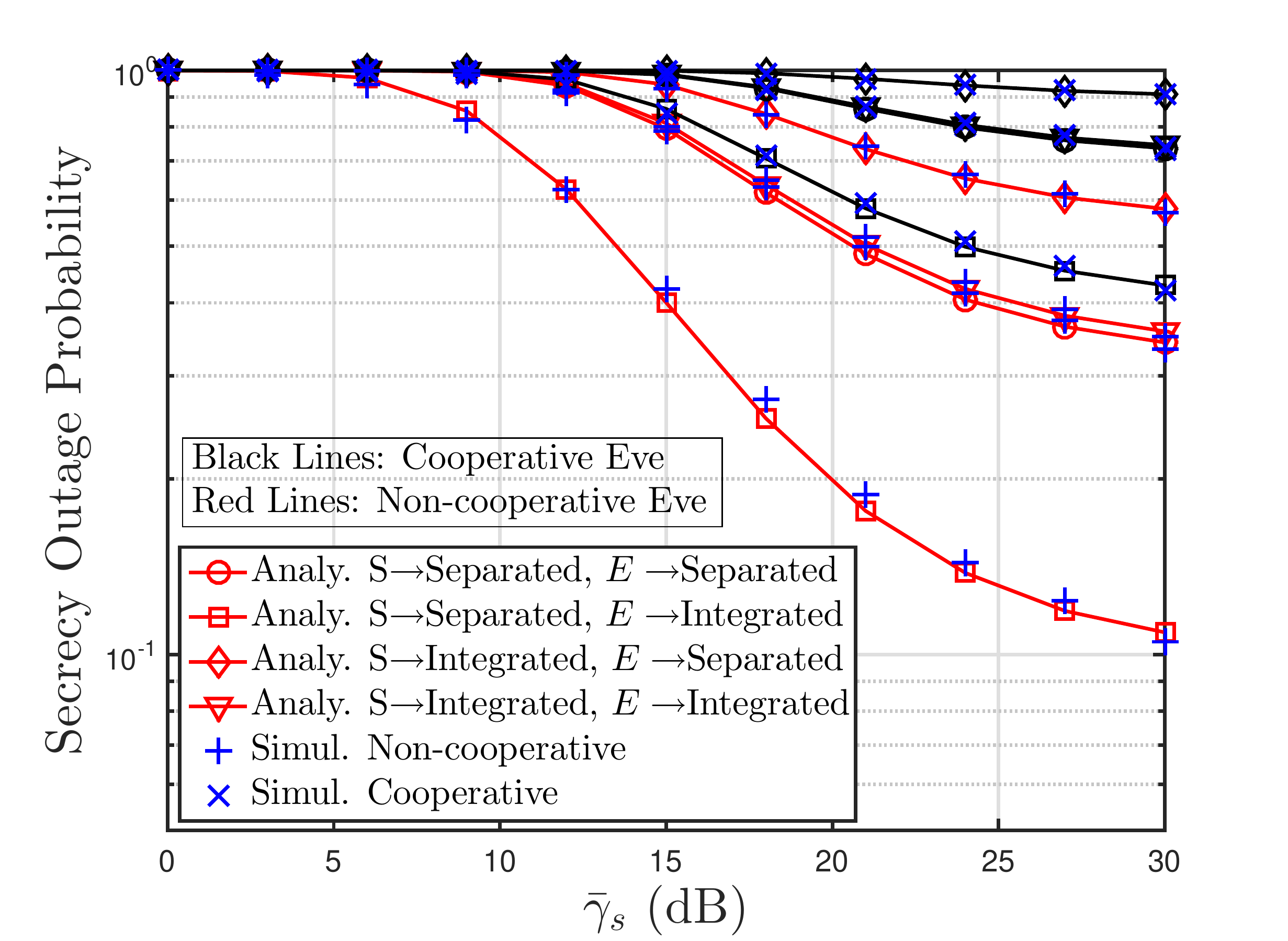}\\
    \centering (a)& \centering (b)
  \end{tabular}%
	\caption{Comparison of secrecy performance between different SWIPT receiver architectures. (a) $R_s=1$ bps/Hz (b) $R_s=2$ bps/Hz.}
	\label{Res_1}
\end{figure*}
\begin{figure*}
  \centering
  \begin{tabular}[htb]{cc}
    \includegraphics[width=.45\textwidth]{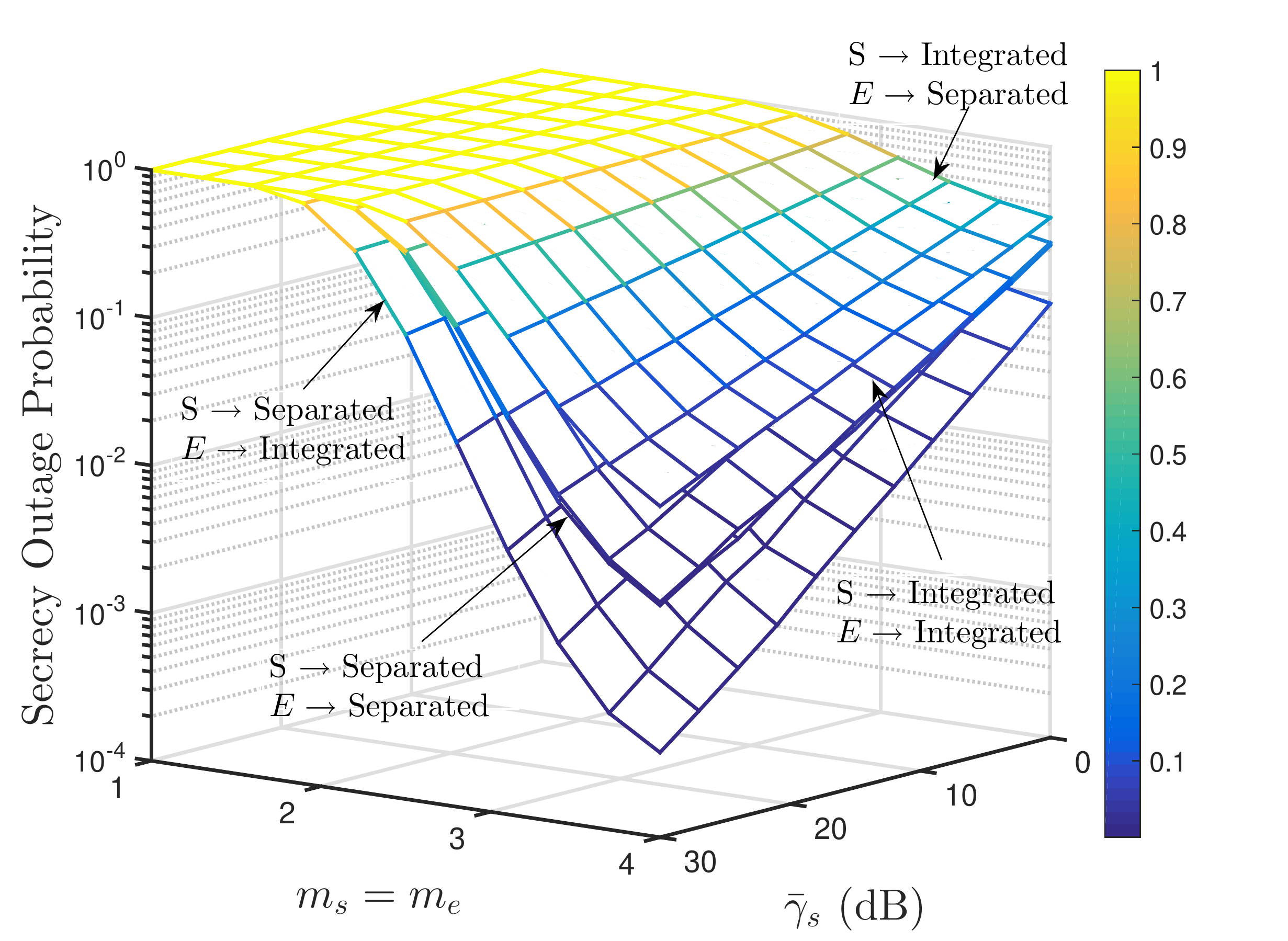}&
    \includegraphics[width=.45\textwidth]{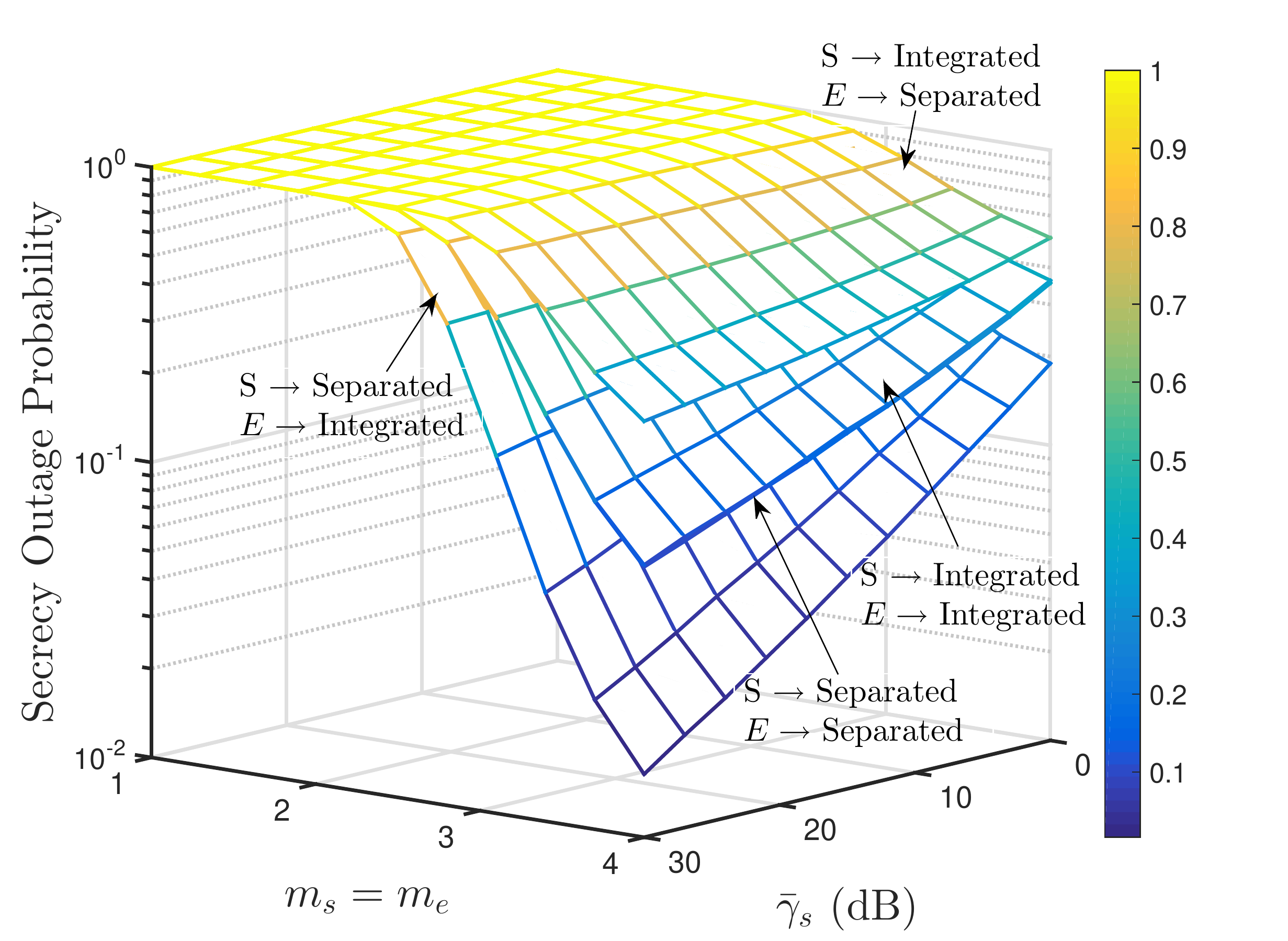}\\
    \centering (a)& \centering (b)
  \end{tabular}%
	\caption{Effect of Nakagami-$m$ parameter and eavesdropper cooperation on secrecy performance, $\delta_s=\delta_e=0.1$. (a) Non-cooperative eavesdroppers (b) Cooperative eavesdroppers.}
	\label{Res_2}
\end{figure*}
\begin{figure}[htb]
\centering
\includegraphics[width=.45\textwidth]{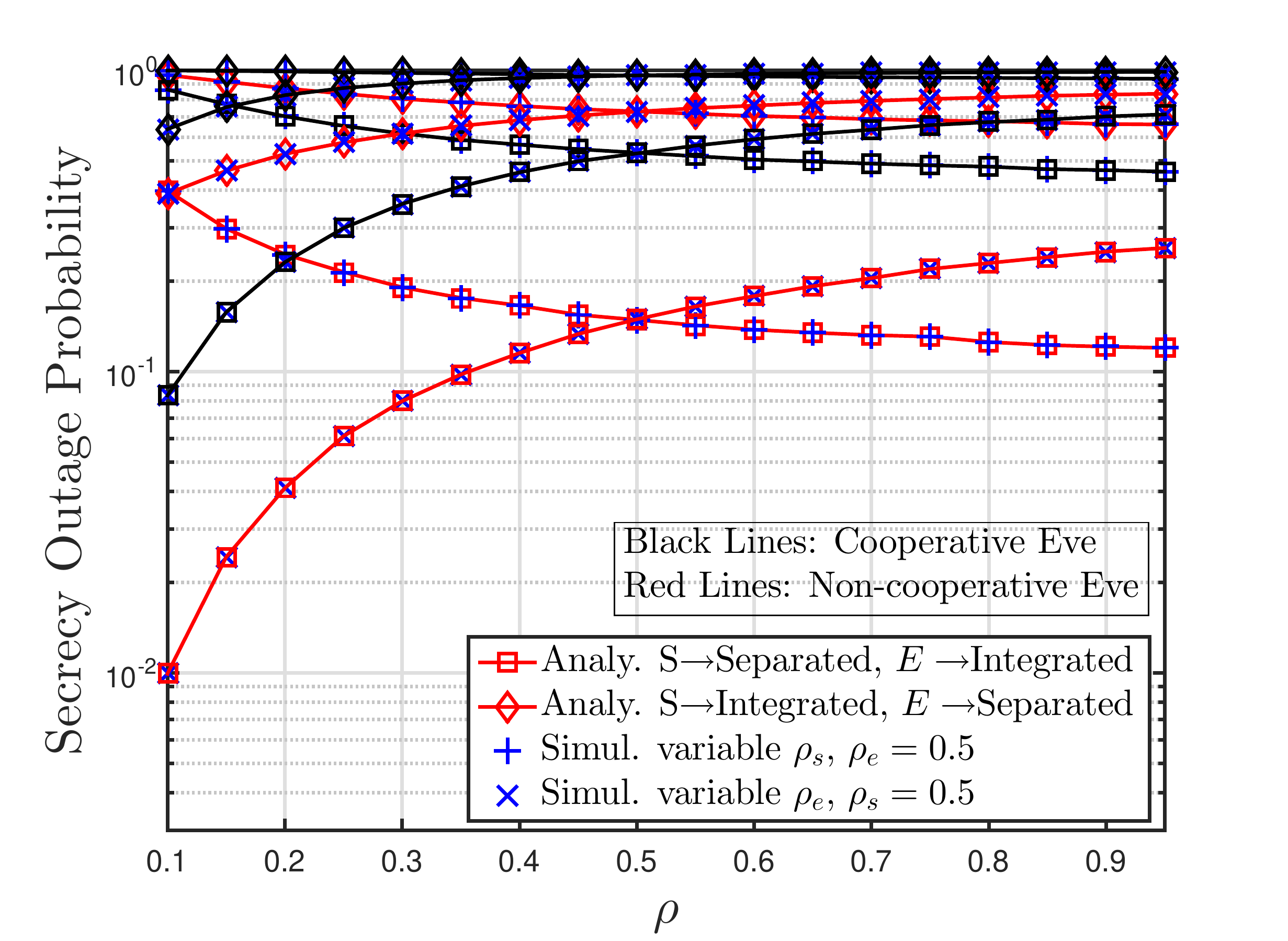}
\caption{Effect of power splitting factor $\rho$ on the secrecy outage probability.}
\label{Res_3}
\end{figure}

Fig. \ref{Res_1} compares the secrecy performance for different combinations of receiver architectures at the legitimate receiver and the eavesdroppers. Fig. \ref{Res_1}(a) shows that for any given value of $\bar{\gamma}_s$, the smallest secrecy outage probability is achieved when S is equipped with a separated and $E$ with an integrated receiver architecture. on the other hand, the secrecy outage probability is the largest for the case when S is equipped with an integrated and $E$ with a separated architecture, all other parameters remaining un-changed. The figure also shows that the outage probability increases with cooperation between the eavesdroppers. Fig. \ref{Res_1}(a) shows that by increasing $\bar{\gamma}_s$ a steady reduction in the outage probability can be achieved. However, at large values ($\bar{\gamma}_s>28$ dB), an outage floor is introduced for both the cooperative and non-cooperative cases, which shows that the outage probability does not decrease despite an increase in the main link SNR. This floor appears because of the channel estimation errors for the main link. By comparing Figs. \ref{Res_1}(a) \& (b), it can be observed that by increasing the target rate $R_s$, for a fixed $\bar{\gamma}_s$, the outage probability increases for all receiver architecture combinations. Finally, comparing the two sub-figures also reveals that the difference between the outage performance with and without eavesdropper cooperation diminishes as $R_s$ is increased from 1 to 2 bps/Hz. All graphs shown in the figures exhibit a good match between the simulation and analytical results, which validates the accuracy of our derived analytical expressions.

Fig. \ref{Res_2} shows the secrecy outage probability surface plotted against $\bar{\gamma}_s$ and the Nakagami-$m$ parameter, for different receiver architectures at S and $E$. Figure \ref{Res_2}(a), for the case of non-cooperative eavesdroppers, shows that the secrecy outage probability decreases with an increase in $m_s=m_e$, which corresponds to a decreasing severity of the channel fading. Moreover, the figure shows that progressively larger values of the Nakagami parameter ($m_s=m_e=m>2$, result in an increasing difference between the secrecy outage probabilities achieved by the 4 receiver combinations; the combination of S separated and $E$ integrated receivers has the smallest outage as already observed in Fig. 3. By comparing Fig. \ref{Res_2}(b), i.e., cooperative eavesdroppers with Fig. \ref{Res_2}(a) for the non-cooperating case, it can be observed that for a given $\gamma_s$ and identical system parameters, cooperation between the eavesdroppers significantly increases the secrecy outage probability relative to that for the non-cooperative case.

Fig. \ref{Res_3} shows the impact of the PS factor $\rho$ on the secrecy outage probability. To separately demonstrate the effect of PS at S only, $\rho_s$ is varied while the PS factor at the eavesdroppers is fixed at $\rho_e=0.5$. Another set of curves shown in Fig. \ref{Res_3} describe the effect of PS at eavesdroppers only, while $\rho_s=0.5$ is maintained for those curves. The figure shows that by increasing values of $\rho_s$ the secrecy outage probability decreases. This is because a larger fraction of the received power is then used for ID at S. In contrast, the secrecy outage probability increases with increasing values of $\rho_e$. This is due to the fact that more power is then allocated by the eavesdroppers to decode the secret message, which diminishes the system's secrecy performance. 
\begin{figure*}
  \centering
  \begin{tabular}[htb]{c}
    \includegraphics[width=.45\textwidth]{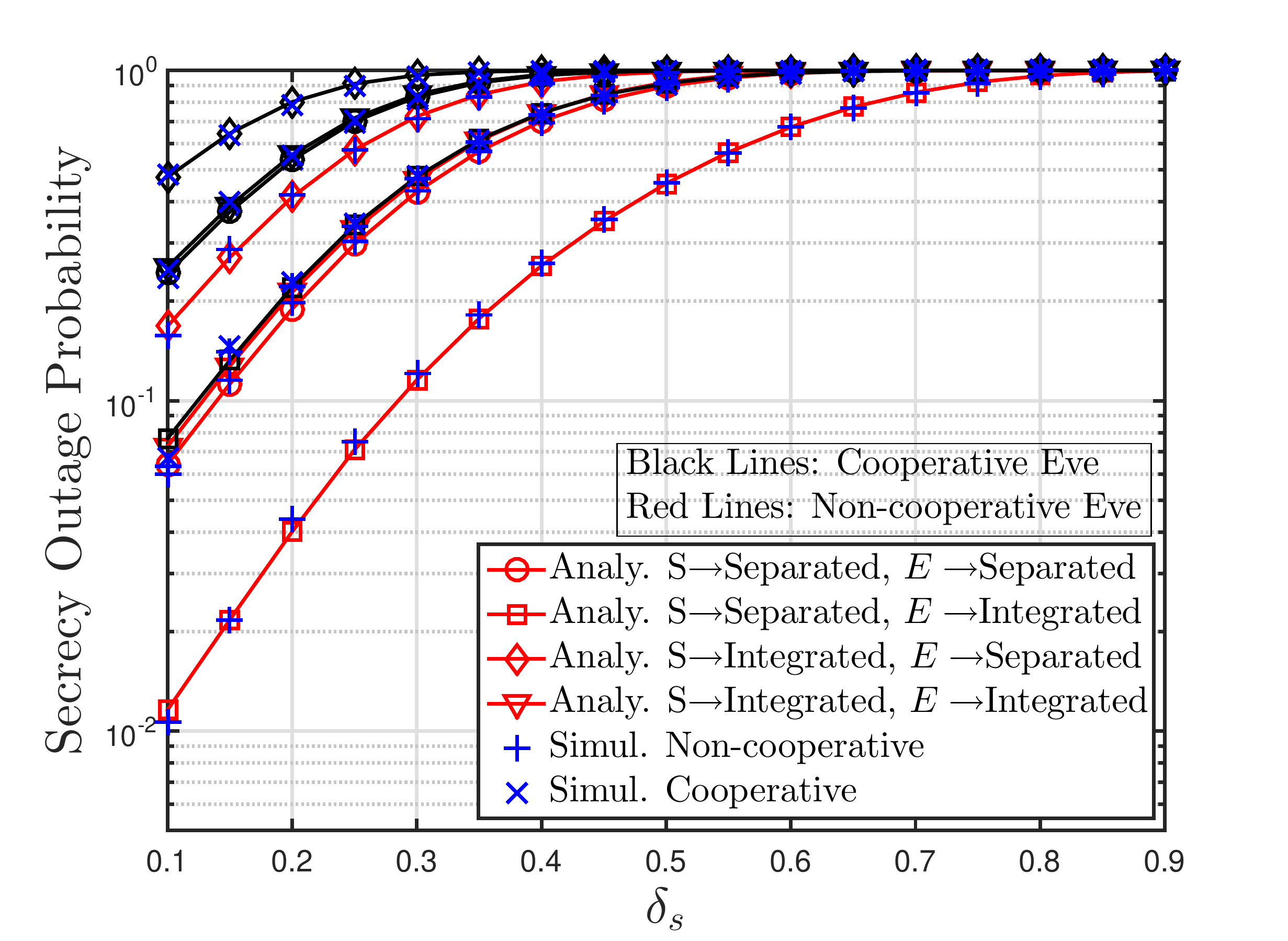}\\
    \small (a)
  \end{tabular}\qquad
	\begin{tabular}[htb]{c}
    \includegraphics[width=.45\textwidth]{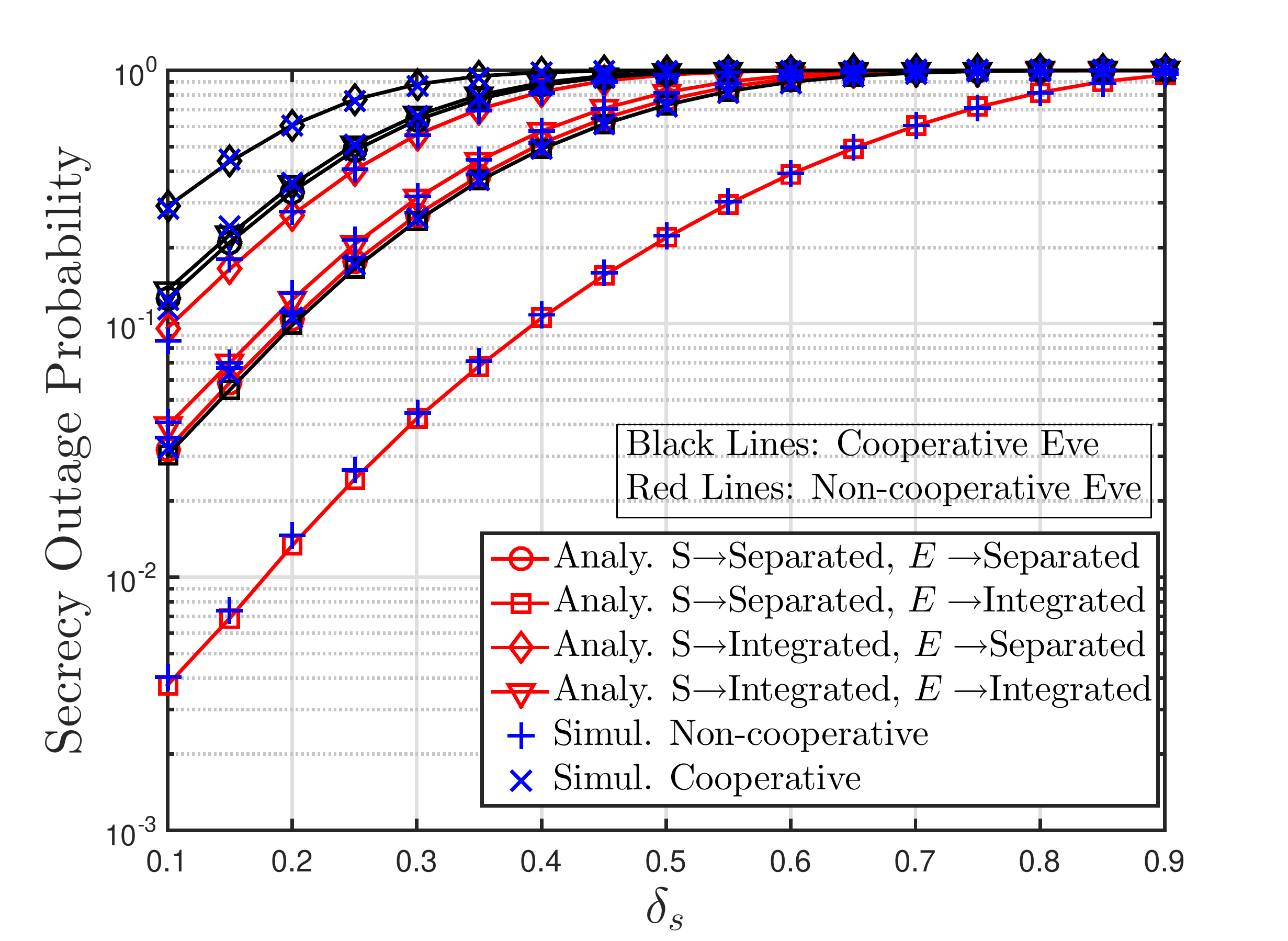}\\
    \small (b)
  \end{tabular}%
	\vspace{0.5cm}
	\centering
  \begin{tabular}[htb]{c}
    \includegraphics[width=.45\textwidth]{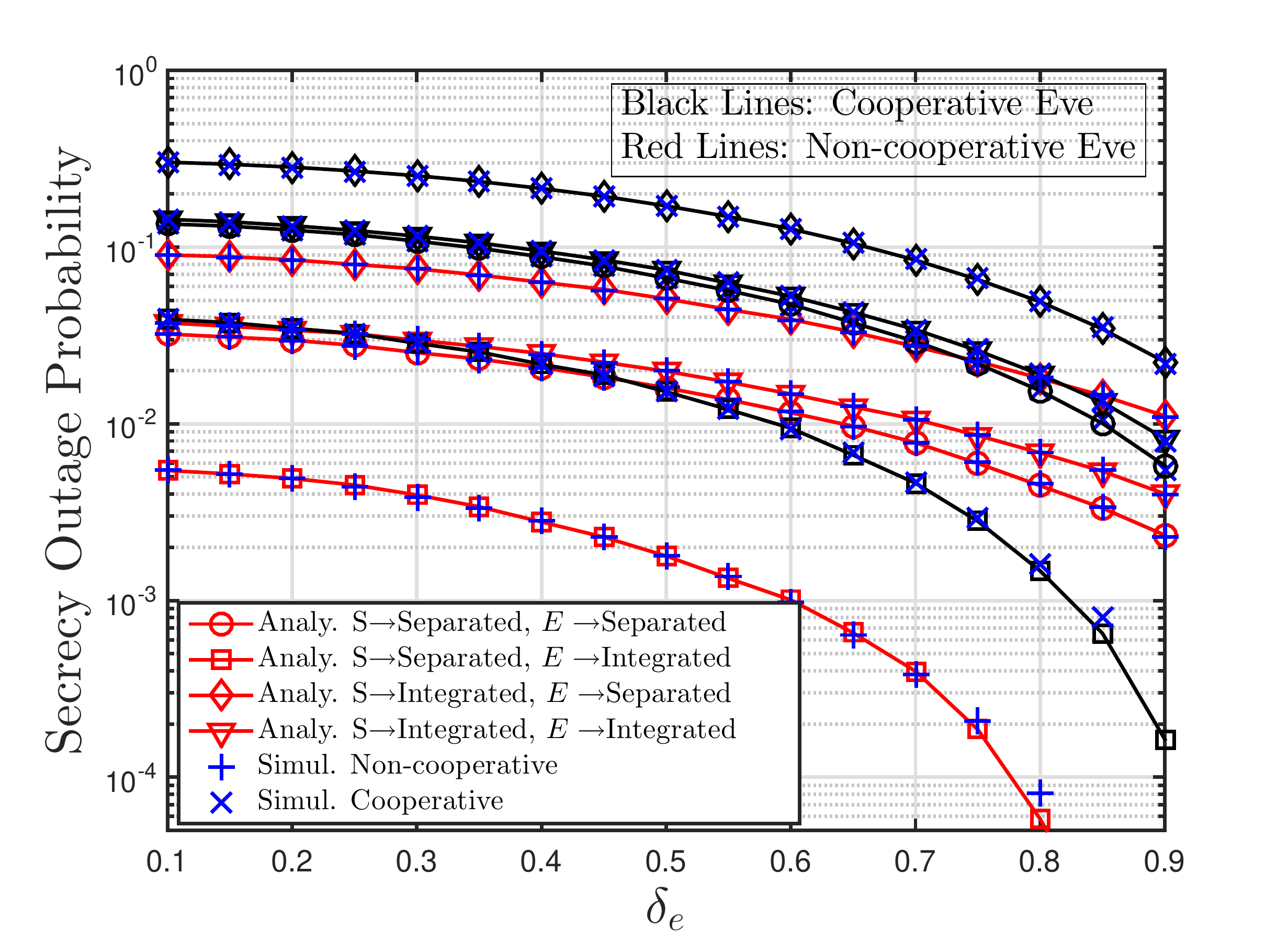}\\
    \small (c)
  \end{tabular}\qquad
	\begin{tabular}[htb]{c}
    \includegraphics[width=.45\textwidth]{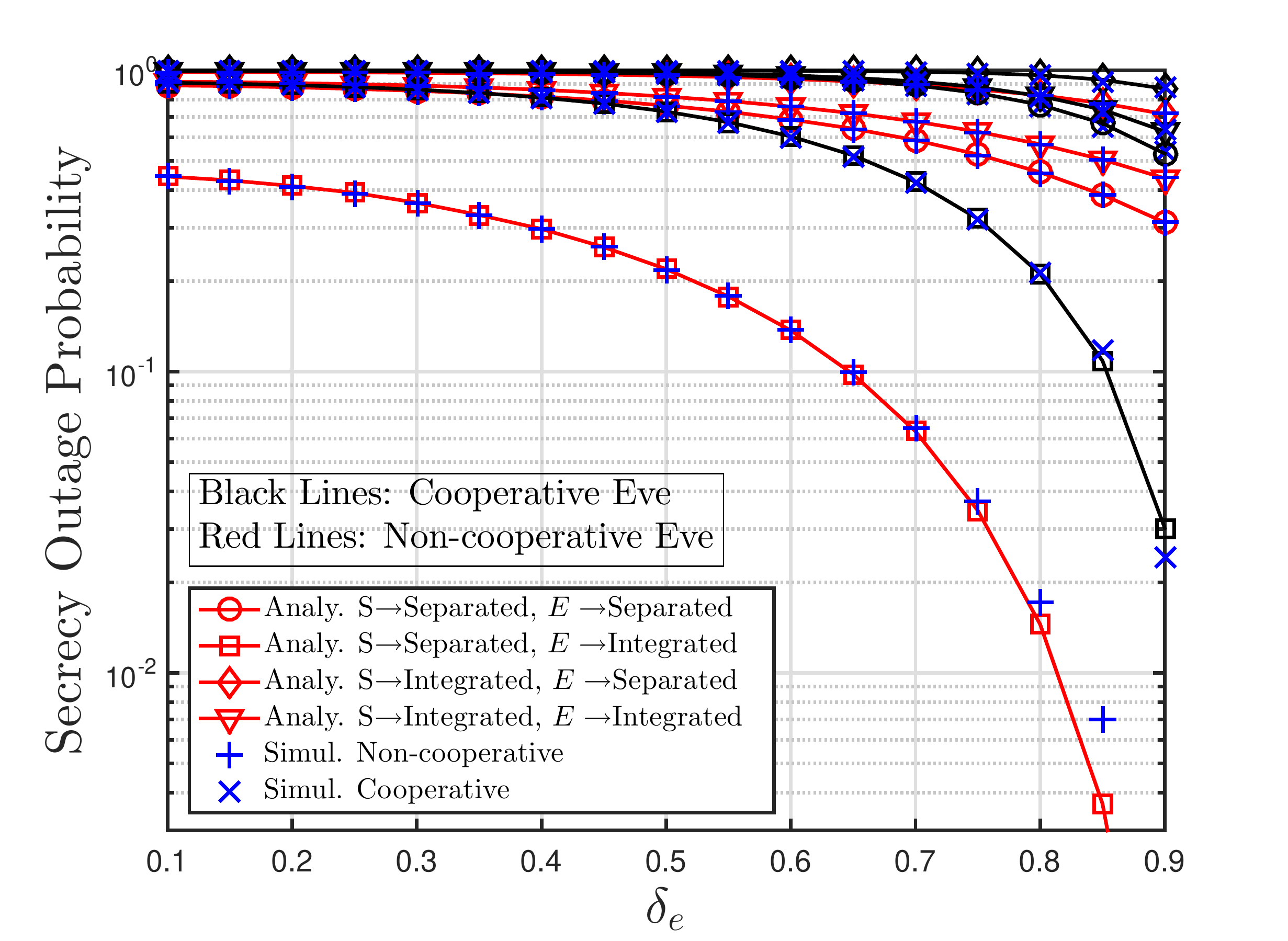}\\
    \small (d)
  \end{tabular}%
	\caption{Effect of imperfect CSI on the secrecy outage probability. (a) variable $\delta_s$, fixed $\delta_e=0.001$. (b) variable $\delta_s$, fixed $\delta_e=0.5$. (c) variable $\delta_e$, fixed $\delta_s=0.001$. (d) variable $\delta_e$, fixed $\delta_s=0.5$.}
	\label{Res_4}
\end{figure*}
\begin{figure}[htb]
\centering
\includegraphics[width=.45\textwidth]{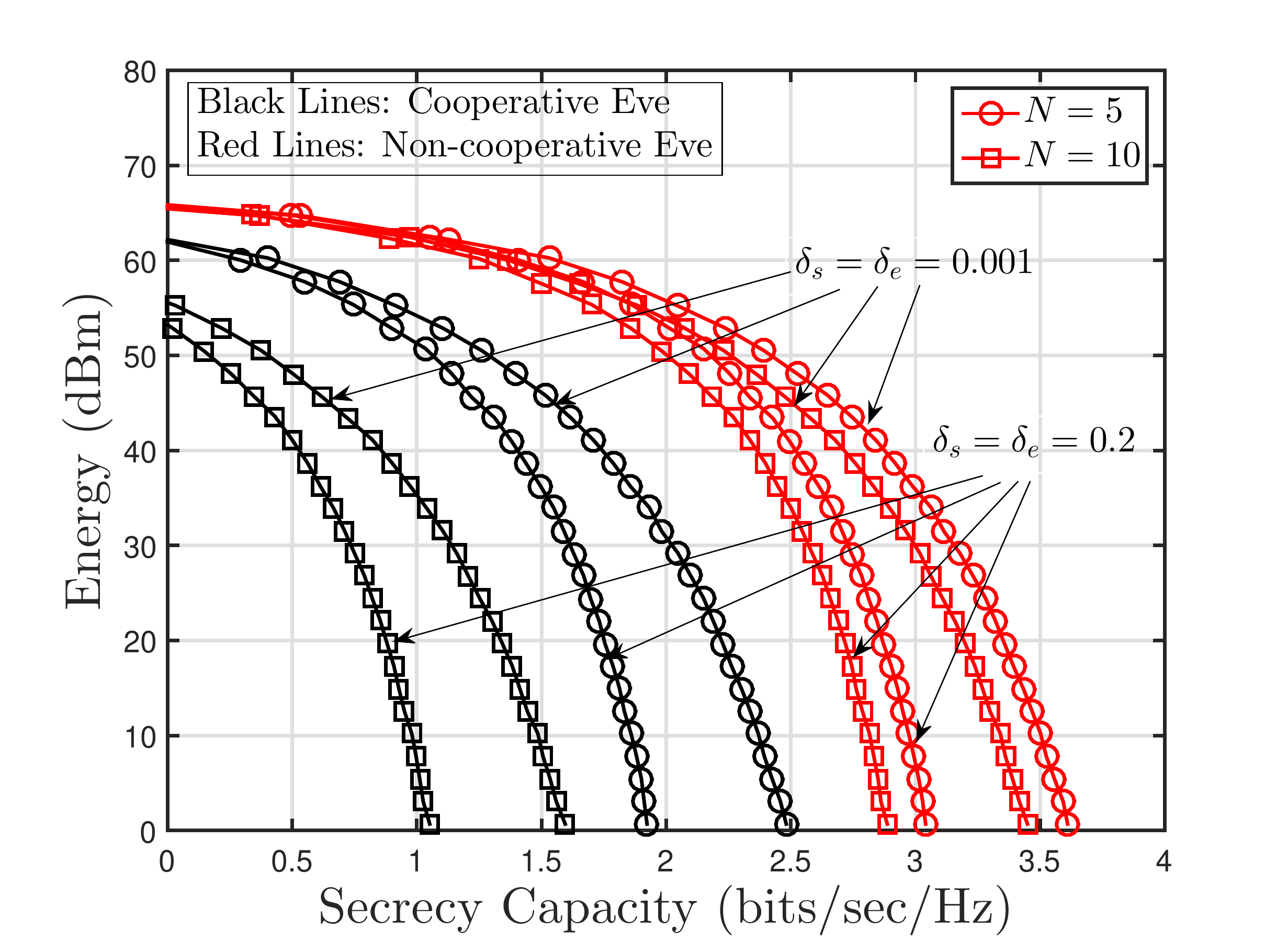}
\caption{Energy-Secrecy Capacity Region. $\rho_s$ varied between 0.01 and 0.99, $\rho_e=0.5$ and $\zeta_e=\zeta_s=0.8$.}
\label{Res_5}
\end{figure}

Fig. \ref{Res_4} shows the impact of the channel estimation errors on the secrecy outage probability. Figs. \ref{Res_4}(a),(b) show that an increase in $\delta_s$, the legitimate receiver's estimation error, degrades the secrecy performance. Whereas, Figs. \ref{Res_4}(c),(d) show that an increase in $\delta_e$, the eavesdropping receiver's estimation error, reduces the secrecy outage probability. This follows from the fact that the secrecy outage event is dependent on the decoding ability of both the legitimate and the eavesdropper nodes. An imperfect channel estimate at the eavesdropper increases its likelihood of incorrect decoding of the secret message, which reduces the information leakage. One may also observe from the figure that an increasing error in CSI estimate of the higher SNR main link has a more dominant effect on the secrecy outage probability than a similar increase in CSI error on the  wiretapping receivers. This can be verified by comparing the relative shift in the secrecy outage curves between Fig. \ref{Res_4}(a) and (b) with the relative shift in the secrecy outage between Fig. \ref{Res_4}(c) and Fig. \ref{Res_4}(d). This effect is more pronounced for the cooperating eavesdroppers case.

Fig. \ref{Res_5} shows the energy-secrecy capacity tradeoff for both cooperative as well as non-cooperative eavesdroppers. Each tradeoff curve is generated by varying $\rho_s$ between 0.01 and 0.99 with fixed $\rho_e=0.5$. However, plotting of each curve in Fig. \ref{Res_5} is restricted to its respective $\rho_s$ sub-interval that produces a non-negative secrecy capacity. This results in different energy levels, harvested according to ($1-\rho_s$), at zero secrecy capacity as shown in Fig. \ref{Res_5}. One may observe from the figure that the enhanced eavesdropper performance due to cooperation diminishes the harvested energy conditional on a non-negative secrecy capacity. The figure also shows that $\delta_s=\delta_e=0.001$ achieves a better energy-secrecy operating point than that of $\delta_s=\delta_e=0.2$, which highlights the significance of having an accurate CSI estimate at the main receiver. Moreover, the figure shows that when the number of eavesdroppers $N$ increases from 5 to 10, the area of the energy-secrecy capacity region decreases significantly for both the cooperative as well as the non-cooperative eavesdroppers. Finally, for a fixed number of eavesdroppers, the energy-secrecy capacity region for non-cooperative eavesdroppers is larger than that of the cooperative eavesdroppers. This highlights the fact that cooperation among the eavesdroppers considerably degrades the secrecy performance of the system.
\section{Conclusion}
This work has investigated the secrecy outage probability for a SWIPT system operating in the presence of cooperative eavesdroppers and different combinations of the SWIPT receiver architectures considered at the legitimate receiver and the eavesdroppers. We derived closed-form expressions for the secrecy outage probability for each of these cases and showed that the smallest secrecy outage probability is achieved when the legitimate receiver has a separated architecture and the eavesdroppers have an integrated SWIPT receiver. The worst-case scenario is when the legitimate receiver has an integrated architecture and the eavesdroppers have separated SWIPT architectures; for a high main link SNR and Nakagami-$m=4$, it was shown that the secrecy outage probabilities achieved for these two extreme cases differ by an order of magnitude. The effect of channel estimation errors was also investigated and it was shown that for the main link average SNR greater than 28 dB, an outage floor appears, i.e., the secrecy outage probability cannot be reduced further due to the channel estimation errors, despite an increase in the main link SNR. Finally, it was shown that cooperation between the eavesdroppers significantly increases the secrecy outage probability relative to that of the non-cooperative case for any combination of receiver architectures. Our results are useful for analyzing the secrecy performance of different SWIPT receiver architectures and eavesdropper cooperation.
\section*{Acknowledgment}
This work is supported by the EU-funded project ATOM-690750, approved under call H2020-MSCA-RISE-2015.
\ifCLASSOPTIONcaptionsoff
  \newpage
\fi



\bibliographystyle{IEEEtran}
\balance
\bibliography{References}
%

\end{document}